\documentclass[12pt]{article}

\title{A Critique of a Polynomial-time SAT Solver Devised by Sergey Gubin}
\author{Ian Christopher \and Dennis Huo \and Bryan Jacobs }

\usepackage{amsmath, theorem, amssymb}
\usepackage{subfigure}
\usepackage{slashbox}
\usepackage{pict2e}
\usepackage{url}

\begin{document}
\maketitle

\begin{abstract}
This paper refutes the validity of the polynomial-time algorithm for solving satisfiability proposed by Sergey Gubin.  Gubin introduces the algorithm using 3-SAT and eventually expands it to accept a broad range of forms of the Boolean satisfiability problem.  Because 3-SAT is NP-complete, the algorithm would have implied P = NP, had it been correct.  Additionally, this paper refutes the correctness of his polynomial-time reduction of SAT to 2-SAT.
\end{abstract}

\newpage

\section{Introduction}
The Boolean satisfiability problem (SAT) deals with determining the satisfiability of a Boolean expression.  That is, given an expression of Boolean literals, ``and'' operators, ``or'' operators, and ``not'' operators, is there an assignment of the literals such that the formula is true?  Within complexity theory, it is a well-known problem that has been heavily researched.  It was one of the first problems to be shown NP-complete \cite{cook}.  Thus, a polynomial-time algorithm to determine the satisfiability of a Boolean expression would imply P = NP.

	Many forms of the Boolean satisfiability problem exist. Despite artificial restrictions, most of these problems remain NP-complete. 3-SAT is one of these special variations. Restrictions are placed on the form of the Boolean expression, but the goal of determining satisfiability remains. 3-SAT expressions are written in conjunctive normal form as such:

\begin{equation}
f = c_1 \wedge c_2 \wedge \ldots \wedge c_m
\label{eq:sat}
\end{equation}

Where $c_k, k=1, 2, \ldots, m$ is a disjunctive clause of three or fewer Boolean literals.  The following is a typical 3-SAT instance:

\begin{equation}
f = (q \vee p \vee r)\wedge(\neg{q} \vee p \vee \neg{r})\wedge(q \vee \neg{p} \vee r)\wedge(\neg{q} \vee \neg{p} \vee \neg{r})
\end{equation}

\section{Gubin's Algorithm}

Sergey Gubin recently proposed a polynomial-time algorithm for solving 3-SAT \cite{3sat}. Following a brief mention of the reducibility of the NP complexity class to this NP-complete problem, a relatively straightforward matrix-trimming algorithm is presented to solve any general 3-SAT problem with $O(n^3)$ time complexity in the number of disjunctive clauses. Furthermore, Gubin reveals that his algorithm can be trivially adapted for general Boolean satisfiability problems in conjunctive normal form. 

Gubin later extends this work to show that his algorithm not only determines the satisfiability of a Boolean formula in conjunctive normal form, but that the particular set of variables satisfying the instance can be recovered from the algorithm \cite{algorithm}. 

In this section, we provide an overview of Gubin's algorithm along with an example of its use. Initially, we are given a SAT instance written as in Equation \ref{eq:sat}.

The fundamental mechanism of the algorithm involves doing pair-wise comparisons of clauses in order to iteratively ``trim'' a set of potential solutions. The problem is first rewritten in a triangular form:

\begin{table}[h]
\begin{tabular}{rrrrr}
\multicolumn{5}{r}{$(c_1 \wedge c_2) \wedge (c_1 \wedge c_3) \wedge (c_1 \wedge c_4) \wedge \ldots \wedge (c_1 \wedge c_m)$} \\
& \multicolumn{4}{r}{$(c_2 \wedge c_3) \wedge (c_2 \wedge c_4) \wedge \ldots \wedge (c_2 \wedge c_m)$} \\
& & \multicolumn{3}{r}{$(c_3 \wedge c_4) \wedge \ldots \wedge (c_3 \wedge c_m)$} \\
& & & & \ldots \\
& & & & $(c_{m-1} \wedge c_m)$
\end{tabular}
\label{tab:triangle}
\end{table}

Each pair of clauses in parentheses is then converted into a matrix. To build such matrices, Gubin introduces the notion of ``compatibility,'' which he ultimately equates to pair-wise satisfiability; the matrix $C_{i,j}$ denotes the ``compatibility matrix'' between clauses $c_i$ and $c_j$, whose boolean entries represent variable assignments satisfying $(c_i \wedge c_j)$. For each matrix $C_{i,j}$, each entry $e_{a,b}$ corresponds to the variable assignments in row $a$ of the truth table for $c_i$ combined with the variable assignments in row $b$ of the truth table for $c_j$. The matrix entry $e_{a,b}$ is true if the following conditions are met:

\begin{enumerate}
\item Row $a$ of $c_i$'s truth table is true, and row $b$ of $c_j$'s truth table is true.
\item Any variables present in both $c_i$ and $c_j$ must not have conflicting truth table assignments in rows $a$, $b$, i.e., if variable $x$ is false in row $a$ of the truth table for $c_i$, $x$ must also be false in row $b$ of the truth table for $c_j$.
\end{enumerate}

After the initial triangular array of matrices is constructed, the algorithm iteratively ``depletes'' each of the matrices in rows $i$ through $m$ with the clause $c_{i-1}$. ``Depletion'' is achieved by eliminating nonzero entries $e_{ab}$ in the matrix $C_{i,j}$ whenever column $a$ of matrix $C_{i-1,i}$ and column $b$ of matrix $C_{i-1,j}$ do not have any nonzero entries in the same row. 

To clarify the notions of ``compatibility'' and ``depletion'', we first present examples of matrix construction and matrix depletion adapted from examples offered by Gubin \cite{examples}. Consider the clauses:

\begin{equation}
c_2 = \neg p \vee q \vee \neg r
\label{eq:c2}
\end{equation}

\begin{equation}
c_3 = p \vee \neg q \vee s
\label{eq:c3}
\end{equation}

The corresponding truth tables for these disjunctive clauses are:

\begin{table}[h]
\centering
\subtable[$c_2$]{
\begin{tabular}{ccc|c}
p & q & r & $\neg p \vee q \vee \neg r$ \\
\hline
0 & 0 & 0 & 1 \\
0 & 0 & 1 & 1 \\
0 & 1 & 0 & 1 \\
0 & 1 & 1 & 1 \\
1 & 0 & 0 & 1 \\
1 & 0 & 1 & 0 \\
1 & 1 & 0 & 1 \\
1 & 1 & 1 & 1 \\
\end{tabular}
}
\subtable[$c_3$]{
\begin{tabular}{ccc|c}
p & q & s & $p \vee \neg q \vee s$ \\
\hline
0 & 0 & 0 & 1 \\
0 & 0 & 1 & 1 \\
0 & 1 & 0 & 0 \\
0 & 1 & 1 & 1 \\
1 & 0 & 0 & 1 \\
1 & 0 & 1 & 1 \\
1 & 1 & 0 & 1 \\
1 & 1 & 1 & 1 \\
\end{tabular}
}
\caption{Truth Tables for Example}
\label{tab:tte}
\end{table}

Since each row denotes a unique combination of variable assignments, an 8x8 matrix can enumerate all possible combinations of variable assignments between two clauses of 3 variables each. The matrix $C_{2,3}$ is constructed by representing the truth table of clause $c_2$ on the vertical axis, and the truth table of clause $c_3$ on the horizontal axis. A matrix entry is determined to be mutually satisfiable if the truth tables for both clauses are true in the rows indexed by the matrix, and if there are no conflicting variable assignments for those variables that are shared (if no variables are shared, then this second condition is automatically satisfied). Applying this to the two clauses shown above results in the following matrix:

\begin{center}
\begin{tabular}{rc}
& $c_3$ \\
$c_2$ & \begin{tabular}{|cccccccc|}
\hline
 1 & 1 & 0 & 0 & 0 & 0 & 0 & 0  \\  
 1 & 1 & 0 & 0 & 0 & 0 & 0 & 0  \\ 
 0 & 0 & 0 & 1 & 0 & 0 & 0 & 0  \\ 
 0 & 0 & 0 & 1 & 0 & 0 & 0 & 0  \\ 
 0 & 0 & 0 & 0 & 1 & 1 & 0 & 0  \\ 
 0 & 0 & 0 & 0 & 0 & 0 & 0 & 0  \\ 
 0 & 0 & 0 & 0 & 0 & 0 & 1 & 1 \\ 
 0 & 0 & 0 & 0 & 0 & 0 & 1 & 1 \\ 
 \hline
\end{tabular}
\end{tabular}
\end{center}

In the above matrix, element (2,2) (row 2, column 2) is true because the second row of the truth table for $c_2$ is ``compatible'' with the second row of $c_3$; both truth table entries are nonzero, and the shared variables $p$ and $q$ can be mutually satisfied by assigning both to false. 

Similarly, (3,3) is false simply because $c_3$ has a zero in row 3 of its truth table. The element (5,4) is not compatible because row 5 of $c_2$ requires $p = 1$, $q = 0$, $r = 0$, while row 4 of $c_3$ requires $p = 0$, $q = 1$, $s = 1$; both $p$ and $q$ cause incompatibilities with this assignment. 

Now, suppose we want to ``deplete'' the above matrix with the clause $c_1 = p \vee  q \vee r$.  This clause has the truth table:

\begin{center}
\begin{tabular}{ccc|c}
p & q & r & $p \vee q \vee r$ \\
\hline
0 & 0 & 0 & 0 \\
0 & 0 & 1 & 1 \\
0 & 1 & 0 & 1 \\
0 & 1 & 1 & 1 \\
1 & 0 & 0 & 1 \\
1 & 0 & 1 & 1 \\
1 & 1 & 0 & 1 \\
1 & 1 & 1 & 1 \\
\end{tabular}
\end{center}

With Gubin's algorithm, we would have computed the compatibility matrices for $C_{1,2}$ and $C_{1,3}$ in the initial step, obtaining the following two matrices:

\begin{center}
\begin{tabular}{rcc}
& $c_2$ & $c_3$ \\
$c_1$ &
\begin{tabular}{|cccccccc|}
\hline
0 & 0 & 0 & 0 & 0 & 0 & 0 & 0  \\ 
0 & 1 & 0 & 0 & 0 & 0 & 0 & 0  \\
0 & 0 & 1 & 0 & 0 & 0 & 0 & 0  \\
 0 & 0 & 0 & 1 & 0 & 0 & 0 & 0  \\
 0 & 0 & 0 & 0 & 1 & 0 & 0 & 0  \\
 0 & 0 & 0 & 0 & 0 & 0 & 0 & 0  \\
 0 & 0 & 0 & 0 & 0 & 0 & 1 & 0 \\ 
 0 & 0 & 0 & 0 & 0 & 0 & 0 & 1 \\ 
 \hline
\end{tabular} &
\begin{tabular}{|cccccccc|}
\hline
 0 & 0 & 0 & 0 & 0 & 0 & 0 & 0  \\  
 1 & 1 & 0 & 0 & 0 & 0 & 0 & 0  \\ 
 0 & 0 & 0 & 1 & 0 & 0 & 0 & 0  \\ 
 0 & 0 & 0 & 1 & 0 & 0 & 0 & 0  \\ 
 0 & 0 & 0 & 0 & 1 & 1 & 0 & 0  \\ 
 0 & 0 & 0 & 0 & 1 & 1 & 0 & 0  \\ 
 0 & 0 & 0 & 0 & 0 & 0 & 1 & 1 \\ 
 0 & 0 & 0 & 0 & 0 & 0 & 1 & 1 \\ 
 \hline
\end{tabular}
\end{tabular}
\end{center}

After depletion, the matrix $C_{2,3}$ becomes:

\begin{center}
\begin{tabular}{rc}
& $c_3$ \\
$c_2$ &
\begin{tabular}{|cccccccc|}
\hline
 0 & 0 & 0 & 0 & 0 & 0 & 0 & 0  \\ 
 1 & 1 & 0 & 0 & 0 & 0 & 0 & 0  \\ 
 0 & 0 & 0 & 1 & 0 & 0 & 0 & 0  \\ 
 0 & 0 & 0 & 1 & 0 & 0 & 0 & 0  \\ 
 0 & 0 & 0 & 0 & 1 & 1 & 0 & 0  \\ 
 0 & 0 & 0 & 0 & 0 & 0 & 0 & 0  \\ 
 0 & 0 & 0 & 0 & 0 & 0 & 1 & 1 \\ 
 0 & 0 & 0 & 0 & 0 & 0 & 1 & 1 \\ 
 \hline
\end{tabular}
\end{tabular}
\end{center}

Comparing to the original matrix, we see that the entries (1,1) and (1,2) have been eliminated. As a specific example, elimination of (1,2) was achieved by comparing column 1 of $C_{1,2}$ to column 2 of $C_{1,3}$, and finding that no nonzero entries existed in the same row of both columns. 

Note here that no additional rows or dimensions are added to the matrix, so we lose the information regarding which particular rows of $C_{1,2}$ and $C_{1,3}$ are compatible with the resulting depleted matrix. 

In the algorithm, all matrices in rows $i$ through $m$ are iteratively ``depleted'' by the clause corresponding to row $i-1$ for $i$ ranging from 2 to $m$. Eventually, once all clauses have contributed to depleting bottom-right matrices, Gubin asserts that the bottom-right matrix corresponds to the entire string of disjunctive clauses. At this point, if there are no matrices with all zeros, Gubin's algorithm determines that problem is satisfiable. Similarly, if at any time there is a matrix containing only zeros, the algorithm determines the problem to be unsatisfiable \cite{3sat}.  

The following is a brief walkthrough of an unsatisfiable 2-SAT example:

\begin{equation}
f = c_1 \wedge c_2 \wedge  c_3 \wedge c_4 = ( \neg p \vee q ) \wedge ( p \vee \neg q ) \wedge ( p \vee q ) \wedge ( \neg p \vee \neg q )
\label{eq:sate}
\end{equation}

Construct the truth tables:

\begin{center}
\begin{tabular}{cccc}
\begin{tabular}{cc|c}
\multicolumn{3}{c}{$(\neg p \vee q)$} \\
$p$ & $q$ & $c_1$ \\
\hline
0 & 0 & 1 \\
0 & 1 & 1 \\
1 & 0 & 0 \\
1 & 1 & 1 \\
\end{tabular}
\begin{tabular}{cc|c}
\multicolumn{3}{c}{$(p \vee \neg q)$} \\
$p$ & $q$ & $c_2$ \\
\hline
0 & 0 & 1 \\
0 & 1 & 0 \\
1 & 0 & 1 \\
1 & 1 & 1 \\
\end{tabular}
\begin{tabular}{cc|c}
\multicolumn{3}{c}{$(p \vee q)$} \\
$p$ & $q$ & $c_3$ \\
\hline
0 & 0 & 0 \\
0 & 1 & 1 \\
1 & 0 & 1 \\
1 & 1 & 1 \\
\end{tabular}
\begin{tabular}{cc|c}
\multicolumn{3}{c}{$(\neg p \vee \neg q)$} \\
$p$ & $q$ & $c_4$ \\
\hline
0 & 0 & 1 \\
0 & 1 & 1 \\
1 & 0 & 1 \\
1 & 1 & 0 \\
\end{tabular}
\end{tabular}
\end{center}

Applying compatibility rules to the truth tables, the following initial triangular array of matrices is created:

\begin{center}
\begin{tabular}{cccc}
& $c_2$ & $c_3$ & $c_4$ \\
$c_1$ &
\begin{tabular}{|cccc|}
\hline
 1 & 0 & 0 & 0 \\
 0 & 0 & 0 & 0 \\
 0 & 0 & 0 & 0 \\
 0 & 0 & 0 & 1 \\
 \hline
\end{tabular} &
\begin{tabular}{|cccc|}
\hline
 0 & 0 & 0 & 0 \\
 0 & 1 & 0 & 0 \\
 0 & 0 & 0 & 0 \\
 0 & 0 & 0 & 1 \\
 \hline
\end{tabular} &
\begin{tabular}{|cccc|}
\hline
 1 & 0 & 0 & 0 \\
 0 & 1 & 0 & 0 \\
 0 & 0 & 0 & 0 \\
 0 & 0 & 0 & 0 \\
 \hline
\end{tabular} \\
&&& \\
$c_2$ & & \begin{tabular}{|cccc|}
\hline
 0 & 0 & 0 & 0 \\
 0 & 0 & 0 & 0 \\
 0 & 0 & 1 & 0 \\
 0 & 0 & 0 & 1 \\
 \hline
\end{tabular} &
\begin{tabular}{|cccc|}
\hline
 1 & 0 & 0 & 0 \\
 0 & 0 & 0 & 0 \\
 0 & 0 & 1 & 0 \\
 0 & 0 & 0 & 0 \\
 \hline
\end{tabular} \\
&&& \\
$c_3$ & & &
\begin{tabular}{|cccc|}
\hline
 0 & 0 & 0 & 0 \\
 0 & 1 & 0 & 0 \\
 0 & 0 & 1 & 0 \\
 0 & 0 & 0 & 0 \\
 \hline
\end{tabular}
 \\
\end{tabular}
\end{center}

The second and third rows of matrices are depleted by $c_1$:

\begin{center}
\begin{tabular}{cccc}
& $c_2$ & $c_3$ & $c_4$ \\
$c_1$ &
\begin{tabular}{|cccc|}
\hline
 1 & 0 & 0 & 0 \\
 0 & 0 & 0 & 0 \\
 0 & 0 & 0 & 0 \\
 0 & 0 & 0 & 1 \\
 \hline
\end{tabular} &
\begin{tabular}{|cccc|}
\hline
 0 & 0 & 0 & 0 \\
 0 & 1 & 0 & 0 \\
 0 & 0 & 0 & 0 \\
 0 & 0 & 0 & 1 \\
 \hline
\end{tabular} &
\begin{tabular}{|cccc|}
\hline
 1 & 0 & 0 & 0 \\
 0 & 1 & 0 & 0 \\
 0 & 0 & 0 & 0 \\
 0 & 0 & 0 & 0 \\
 \hline
\end{tabular} \\
&&& \\
$c_2$ & & \begin{tabular}{|cccc|}
\hline
 0 & 0 & 0 & 0 \\
 0 & 0 & 0 & 0 \\
 0 & 0 & 0 & 0 \\
 0 & 0 & 0 & 1 \\
 \hline
\end{tabular} &
\begin{tabular}{|cccc|}
\hline
 1 & 0 & 0 & 0 \\
 0 & 0 & 0 & 0 \\
 0 & 0 & 0 & 0 \\
 0 & 0 & 0 & 0 \\
 \hline
\end{tabular} \\
&&& \\
$c_3$ & & &
\begin{tabular}{|cccc|}
\hline
 0 & 0 & 0 & 0 \\
 0 & 1 & 0 & 0 \\
 0 & 0 & 0 & 0 \\
 0 & 0 & 0 & 0 \\
 \hline
\end{tabular} \\
\end{tabular}
\end{center}

Finally, the third row is depleted with $c_2$: 

\begin{center}
\begin{tabular}{cccc}
& $c_2$ & $c_3$ & $c_4$ \\
$c_1$ &
\begin{tabular}{|cccc|}
 \hline
 1 & 0 & 0 & 0 \\
 0 & 0 & 0 & 0 \\
 0 & 0 & 0 & 0 \\
 0 & 0 & 0 & 1 \\
 \hline
\end{tabular} &
\begin{tabular}{|cccc|}
 \hline
 0 & 0 & 0 & 0 \\
 0 & 1 & 0 & 0 \\
 0 & 0 & 0 & 0 \\
 0 & 0 & 0 & 1 \\
 \hline
\end{tabular} &
\begin{tabular}{|cccc|}
\hline
 1 & 0 & 0 & 0 \\
 0 & 1 & 0 & 0 \\
 0 & 0 & 0 & 0 \\
 0 & 0 & 0 & 0 \\
 \hline
\end{tabular} \\
& & & \\
$c_2$ & & \begin{tabular}{|cccc|}
 \hline
 0 & 0 & 0 & 0 \\
 0 & 0 & 0 & 0 \\
 0 & 0 & 0 & 0 \\
 0 & 0 & 0 & 1 \\
 \hline
\end{tabular} &
\begin{tabular}{|cccc|}
\hline
1 & 0 & 0 & 0 \\
0 & 0 & 0 & 0 \\
0 & 0 & 0 & 0 \\
0 & 0 & 0 & 0 \\
\hline
\end{tabular} \\
& & & \\
$c_3$ & & & \begin{tabular}{|cccc|}
\hline
 0 & 0 & 0 & 0 \\
 0 & 0 & 0 & 0 \\
 0 & 0 & 0 & 0 \\
 0 & 0 & 0 & 0 \\
\hline
\end{tabular} \\
\end{tabular}
\end{center}

Since the bottom-right matrix is now all zeros, the algorithm has successfully shown that this 2-SAT problem is unsatisfiable. The same method is used for 3-SAT or general SAT problems, where each matrix will simply have varying dimensions depending on the size of each disjunctive clause. For more examples, refer to Gubin \cite{examples}.

\section{Comments and Analysis}
Gubin's analysis of asymptotic complexity is correct, in that the algorithm requires O($m^{3}$) matrix compatibility comparisons/depletions. Considering the relatively sparse matrices, an efficient implementation of the algorithm could even be developed.  Unfortunately, the algorithm has significant shortfalls in other areas. 

An immediate sense of information loss emerges after reading the algorithm. Before depletion, the algorithm seems to work well, but during depletion, problems relating to this loss arise.  During each iteration of depletion, Gubin's algorithm could be comparing two compatibility matrices that share just three literals. So, the sixty four bit  compatibility matrix could be forced to store information on nine distinct literals.  Informally speaking, this is not enough space to store adequate information on accepting assignments for this many literals.  There are $2^{9}$ possible combinations of these literals, which is far too much data to store in an eight by eight matrix.  Modifying the size of the depleted matrix (by adding additional dimensions) could fix the algorithm, but it would also require an exponential amount of time.  Regardless, this lost information will catch up to the algorithm and cause errors.

Beyond this, the entire algorithm relies on the correctness of the notion of ``compatibility'' and ``depletion''; the question arises of whether a nonzero entry in a depleted compatibility matrix does, in fact, correspond to a solution to an entire string of mutually satisfiable disjunctive clauses. 

Another prominent issue is the fact that, after depleting a matrix with a particular clause, no information is recorded as to which precise rows of the truth table for that clause are the ones mutually compatible with the entire matrix. Given that initial matrices require 64 (8x8) bits to store all configurations of up to six literals between two 3-SAT clauses, it seems counterintuitive that no extra space is needed to store the $2^9$ interdependent configurations of three clauses, $2^{12}$ configurations of four clauses, or ultimately, $2^m$ configurations over the entire 3-SAT string. On the contrary, the algorithm computes the column-matchup corresponding to a valid configuration in each depletion stage, then \emph{discards} this information at precisely the moment a brute-force method would require \emph{more} information.   

Consider now a 3-SAT case with four disjunctive clauses. Suppose any three are satisfiable without the fourth clause, but that overall, the formula is not satisfiable. Since the algorithm is designed such that clause $c_2$ is never used to deplete matrices built from $c_1$, clause $c_3$ is never used to deplete matrices built from $c_2$, etc, the only matrix that will eventually consider all four clauses is the bottom-right matrix, denoted $C_{3,4}$. 

Suppose $c_3$ and $c_4$ have compatible entries, and that depletion by $c_1$ does not eliminate any entries from this matrix. However, those variable assignments for $c_1$ which are able to match each of the entries in matrix $C_{3,4}$ could very well be precisely those entries which require $c_2$ to be false. If such a case exists, we immediately see that after depletion by $c_1$, we discard the information regarding the requirements imposed by $c_1$, and so the algorithm goes on to incorrectly predict that the problem is indeed satisfiable.

Does such a 3-SAT instance exist?  Yes, for an almost trivial case:

\begin{equation}
f = (a \vee b \vee c) \wedge (\neg a) \wedge (\neg b) \wedge (\neg c)
\end{equation}

Obviously, this is unsatisfiable.  One of the three variables must be true, but a clause forbidding each one to hold a true value is present.  This satisfies the above
assertion that any triplet of the clauses is satisfiable.

Let's walk through Gubin's algorithm by hand.  Truth tables:

\begin{center}
\begin{tabular}{cccc}
\begin{tabular}{ccc|c}
a & b & c & $c_1$ \\
\hline
0 & 0 & 0 & 0 \\
0 & 0 & 1 & 1 \\
0 & 1 & 0 & 1 \\
1 & 0 & 0 & 1 \\
0 & 1 & 1 & 1 \\
1 & 0 & 1 & 1 \\
1 & 1 & 0 & 1 \\
1 & 1 & 1 & 1 \\
\end{tabular} &
\begin{tabular}{c|c}
a & $c_2$ \\
\hline
0 & 1 \\
1 & 0 \\
\end{tabular} &
\begin{tabular}{c|c}
b & $c_3$ \\
\hline
0 & 1 \\
1 & 0 \\
\end{tabular} &
\begin{tabular}{c|c}
c & $c_4$ \\
\hline
0 & 1 \\
1 & 0 \\
\end{tabular}
\end{tabular}
\end{center}

Compatibility matrices:

\begin{center}
\begin{tabular}{cccc}
& $c_2$ & $c_3$ & $c_4$ \\
$c_1$ & 
\begin{tabular}{|cc|}
\hline
0 & 0 \\
1 & 0 \\
1 & 0 \\
0 & 0 \\
1 & 0 \\
0 & 0 \\
0 & 0 \\
0 & 0 \\
\hline
\end{tabular}
 &
\begin{tabular}{|cc|}
\hline
0 & 0 \\
1 & 0 \\
0 & 0 \\
1 & 0 \\
0 & 0 \\
1 & 0 \\
0 & 0 \\
0 & 0 \\
\hline
\end{tabular}
 &
\begin{tabular}{|cc|}
\hline
0 & 0 \\
0 & 0 \\
1 & 0 \\
1 & 0 \\
0 & 0 \\
0 & 0 \\
1 & 0 \\
0 & 0 \\
\hline
\end{tabular}
 \\
 && \\
$c_2$ & & \begin{tabular}{|cc|}
\hline
1 & 0 \\
0 & 0 \\
\hline
\end{tabular} &
\begin{tabular}{|cc|}
\hline
1 & 0 \\
0 & 0 \\
\hline
\end{tabular} \\
&& \\
$c_3$ & & &
\begin{tabular}{|cc|}
\hline
1 & 0 \\
0 & 0 \\
\hline
\end{tabular} \\
\end{tabular}
\end{center}

During the first depletion step, the matrices remain unchanged; the last three clauses have no common variables,
and the $C_{1,2}$ matrix causes no change in any matrix in row two or below.  During the second and the final depletion steps, the same is true.  So, no pattern
of unsatisfiability arises.

Interestingly, changing the order in which the clauses are specified causes this problem to not occur.  If $c_1$ were specified last, then depleting on $c_2$ would
eliminate two rows, and depleting on $c_3$ would eliminate the remaining one.  $C_{1,4}$ would have no entries left and the algorithm's result would be correct.  However,
just trying all possible orderings would require exponential time.  The fact that the bottom-right matrix does not contain
complete information about all clauses (it cannot, because conjunction is transitive and we have just shown that compatibility via Gubin's method is not)
directly contradicts Gubin's assertion that it is logically equivalent to the conjunction of all the formula's clauses.

\section{A Previous Counterexample}
Hegerle presents a purported counterexample to Gubin's algorithm \cite{counterexample}.  Let
$B$ be the set of boolean variables in the original 3-SAT instance. Let $C$ be the set of clauses in that
instance.  Hegerle asserts that Gubin's algorithm reduces to three patterns:

\begin{enumerate}
\item $\exists \alpha \in B$ such that $\{\alpha, \neg \alpha\} \subseteq C$
\item $\exists \alpha, \beta \in B$  such that $\alpha \ne \beta$ and $\{\alpha \vee \beta, \alpha \vee \neg \beta, \neg \alpha \vee \beta, \neg \alpha \vee \neg \beta\} \subseteq C$
\item $\exists \alpha, \beta, \gamma \in B$  such that
$\alpha \ne \beta \ne \gamma$ and $\{\alpha \vee \beta \vee \gamma, \alpha \vee \beta \vee \neg \gamma, \alpha \vee \neg \beta \vee \gamma, \alpha \vee \neg \beta \vee \neg \gamma,
\neg \alpha \vee \beta \vee \gamma, \neg \alpha \vee \beta \vee \neg \gamma, \neg \alpha \vee \neg \beta \vee \gamma, \neg \alpha \vee \neg \beta \vee \neg \gamma\} \subseteq C$
\end{enumerate}

Hegerle claims that Gubin's algorithm will deem a formula unsatisfiable if and only if at least one of the above is true.

This claim is false.  While it is correct that Gubin's algorithm only simultaneously considers three clauses (two in a compatibility matrix
and one on which that matrix is being depleted), some information about other clauses is passed through the depletion process.  As a result,
Gubin's algorithm can handle some formulas which do not contain all possible clauses of a given length $k$, $k \leq 3$.

Indeed, Gubin himself demonstrates how his algorithm functions in this case \cite{examples}.

\section{Addressing a Reduction of SAT to 2-SAT}
Gubin also presents a polynomial-time reduction of SAT to 2-SAT \cite{reduction}.  In this reduction, he first converts the 3-SAT instance to
the form $g \wedge h = true$, where $g$ is a conjunctive combination of clauses and $h$ is a conjunctive subset of the clauses therein.  So,
$g \wedge h = true$ may be viewed as a SAT instance.
For brevity, let us ignore the method he employs to perform this conversion, as it is not where the flaw presented herein lies.

After producing $g$ and $h$, Gubin applies the matrix-compatibility method described in his previous paper \cite{3sat} to this new equation.
He asserts that the bottom-right compatibility matrix after the depletion steps is all zeros if and only if the original formula is unsatisfiable.
However, we have shown this to not always be the case.

When the bottom-right matrix contains nonzero entries, Gubin then enumerates it such that each element is labeled $y_i$. 
Finally, he creates a set of new clauses $\eta_i$, each containing solely the literal $b_i$ if $y_i$ 1, or $\neg b_i$ if $y_i$ is 0.
This is a 1-SAT instance.

Now, back to our objection.  We have already shown that the final matrix after depletion may contain nonzero entries even when the 3-SAT instance cannot
be satisfied.  If this is the case, Gubin will generate a 1-SAT instance where each variable is only contained in a single clause, and only once within
that clause.  Such a problem is always satisfiable: just assign a variable to be false if its one occurrence is negated, and true otherwise.  The
original 3-SAT problem was not satisfiable.  Therefore, the reduction is not correct.

It is possible that the way Gubin converts the 3-SAT instance to $g \wedge h$ eliminates all cases where his matrix-compatibility algorithm fails.  However, no
proof of this is presented within his paper \cite{reduction}.  The reduction, as it stands, is not correct.
\section{Conclusion}
While Gubin's algorithm does take polynomial time, it does not accurately determine satisfiability.  Its underlying principles are reasonable, but the relaxation of the problem at various stages dooms its validity.  Keeping track of all possible Boolean assignments that lead to satisfiability at each stage would fix the algorithm.  Unfortunately, this would require an exponential amount of time when using a method similiar to the one presented in the original algorithm.  Gubin's algorithm fails to prove that P = NP. This remains one of the larger unanswered questions in theoretical computer science.
\section{Acknowledgements}
Special thanks to Stan Park and Lane Hemaspaandra for their insightful comments and suggestions
during the drafting process. Their input helped us deliver as thorough of a discussion
as possible.

\newpage

\bibliography{report}

\end{document}